\documentclass[aps,prl,twocolumn,floatfix,showpacs,groupedaddress]{revtex4}

\usepackage{amsmath}
\usepackage{graphicx}

\newcommand{\myvec}[1]{{\mathbf{#1}}}
\newcommand{\iis}{_{i\sigma}}

\newcommand{\is}{_{\sigma}}
\newcommand{\iu}{_{\uparrow}}
\newcommand{\id}{_{\downarrow}}
\newcommand{\iks}{_{k\sigma}}
\newcommand{\iku}{_{k\uparrow}}
\newcommand{\ikd}{_{k\downarrow}}
\newcommand{\ikmqd}{_{k-q\downarrow}}
\newcommand{\ikpqu}{_{k+q\uparrow}}
\newcommand{\pd}{^{\phantom{\dagger}}}
\newcommand{\iL}{_{\text{L}}}
\newcommand{\iR}{_{\text{R}}}
\newcommand{\hL}{^{\text{L}}}
\newcommand{\hR}{^{\text{R}}}
\newcommand{\ret}{^{\text{R}}}
\newcommand{\euone}{\widetilde\varepsilon\iu}
\newcommand{\eutwo}{\widetilde\varepsilon'\iu}
\newcommand{\edone}{\widetilde\varepsilon\id}

\newcommand{\sect}[1]{\textit{#1.}---}
\begin{document}

\title{Fingerprints of the Magnetic Polaron in Nonequilibrium Electron
  Transport through a Quantum Wire Coupled to a Ferromagnetic Spin
  Chain}
\author{Frank Reininghaus}
\author{Thomas Korb}
\author{Herbert Schoeller}
\affiliation{Institut f\"ur Theoretische Physik A,
  RWTH Aachen, 52056 Aachen, Germany}
\date{\today}

\begin{abstract}
We study nonequilibrium quantum transport through a mesoscopic wire
coupled via local exchange to a ferromagnetic spin chain. Using the
Keldysh formalism in the self-consistent Born approximation, we 
identify fingerprints of the magnetic polaron state formed
by hybridization of electronic and magnon states. Because of its
low decoherence rate, we find coherent transport signals. Both
elastic and inelastic peaks of the differential conductance are
discussed as a function of external magnetic fields, the polarization
of the leads and the electronic level spacing of the wire.
\end{abstract}


\maketitle

\sect{Introduction}In recent years, the field of Spintronics has
attracted increasing
interest~\cite{WolfSpintronicsReview,ZuticSpintronicsReview}. A
considerable amount of theoretical and experimental attention has been
focused on transport phenomena, especially spin-dependent charge
currents in low-dimensional structures made of magnetic
materials~\cite{Rodrigues_MagneticNanowires,
  Avishai,GouldMolenkamp_SelfAssembledQD},
but also transport of magnetization through insulating spin chains and
quantum dots~\cite{MeierLoss,Strelcyk}. 

In this Letter, we study the interplay between nonequilibrium electron
transport and magnetic degrees of freedom in a one-dimensional
system. The model under consideration is a finite quantum wire which
is coupled via local exchange to a one-dimensional ferromagnetic
Heisenberg spin chain and via tunneling to two large 
electronic reservoirs. Examples of one-dimensional systems which
exhibit ferromagnetic coupling of localized spins are so-called
``sandwich clusters'' formed from vanadium and
benzene~\cite{V_Benzene_Sandwich_MiyajimaNakajima,
  V_Benzene_Sandwich_Wang,V_Benzene_Sandwich_Kang}.
Usually, one would expect that emission of magnons in the spin chain
will lead to a relaxation and dephasing of the electron spins
antiparallel to the spin direction of the spin chain, leading to
incoherent transport for this spin direction. However, it was shown in
several works on ferromagnetic semiconductors
\cite{MethfesselMattis,Richmond,ShastryMattis,NoltingReview} that a
single electron with antiparallel spin direction to 
the localized spins can hybridize with one-magnon states to form the
so-called \textit{magnetic polaron} states. These states form a band
which is separated from the band of scattering states, and therefore
have a low decoherence rate. The aim of this Letter is to
find fingerprints of these states in coherent transport signals at low
temperatures by studying the differential conductance as a function of
bias voltage. We note that one-electron scattering in finite
quantum spin chains has been studied in Ref.~\cite{Avishai} at low
temperatures with the result of an interesting resonance structure as
a function of the Fermi level. However, this work was restricted to
linear transport, so that only states near the Fermi level contributed
to transport. Therefore, the influence of the magnetic polaron states
(lying outside the band of scattering states) was not probed there.

We calculate the differential conductance $G=\frac{dI}{dV}$ for large
($N=1000$ sites) and small ($N=12$ sites) systems which differ in the
electronic level spacing. We find peak structures which are due to
elastic and inelastic transport processes. The
applied magnetic field, the spin polarizations of the leads and the
bias voltage affect the energies and decay rates of the electronic
states of the system. One can thus control the position and height of
the peaks in the differential conductance and identify the processes
which contribute to the current.  

\begin{figure}[b]
  \includegraphics[width=\linewidth]{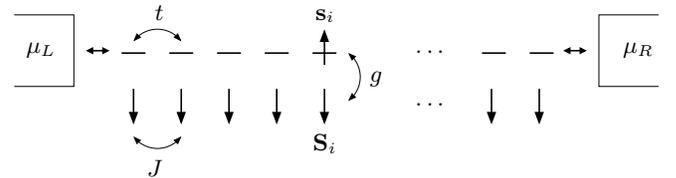}
  \caption{
    \label{fig:model}
    The model under consideration. The conduction electrons can hop
    between neighboring sites in the wire, and the localized spins are
    coupled ferromagnetically. There is a local coupling between the
    spin of the conduction electrons and the localized spin at each
    site. The wire is coupled to two leads with the chemical potential
    $\mu\iL$ and $\mu\iR$, respectively.}
\end{figure}

\sect{Hamiltonian}We employ a tight-binding model for the quantum wire
and a Heisenberg model with ferromagnetic coupling $J>0$ for the spin
chain (see Fig.~\ref{fig:model}). They have the same lattice constant $a$ and Zeeman splittings
$h_e$ and $E_Z$, respectively: 
\begin{align*}
  H_{\text{wire}}&=-t\sum_{i\sigma}\left(c^\dagger\iis
    c\pd_{i+1\sigma}+\text{H.c.}\right)+\frac12
  h_e\sum_{i\sigma}\sigma 
  c^\dagger\iis c\pd\iis\\
  H_{\text{spin}}&=-J\sum_i\myvec{S}_i\myvec{S}_{i+1}+E_Z\sum_i S^z_i
\end{align*}
with $\hbar=1$, $\sigma=\pm$ (we will frequently write $\sigma=\uparrow$ or
$\sigma=\downarrow$ instead). We consider low electron densities in
the wire and therefore neglect the Coulomb interaction.

We follow~\cite{Richmond} and use the Holstein-Primakoff
transformation (HPT)~\cite{HolsteinPrimakoff} to replace the spin operators
in the chain by boson operators: 
$S^+_i\approx\sqrt{2S}\,b^\dagger_i$, $S^-_i\approx\sqrt{2S}\,b_i$,
$S^z_i=b^\dagger_i b\pd_i-S$. This approximation is valid if the spin
chain is near its ferromagnetic ground state where $\langle
S^z_i\rangle=-S$. The Zeeman energy $E_Z$ must be sufficiently large
to ensure that this is the case. Using periodic boundary conditions,
we can write 
$H_{\text{wire}}=\sum_{k\sigma}\varepsilon\pd\iks c^\dagger\iks c\pd\iks$,
$H_{\text{spin}}=E_0+\sum_k \omega\pd_k b^\dagger_k b\pd_k$,
where $E_0=-NJS^2-NSE_Z$ is the ground state energy of the spin chain,
$b^\dagger_k$ and $b\pd_k$ are creation and annihilation operators for
magnons, and $\varepsilon\iks=-2t\cos(ka)+\frac12 \sigma h_e$,
$\omega_k=2JS(1-\cos(ka))+E_Z$ are one-electron and one-magnon
energies, respectively. If $E_Z\gg JS$, we can assume that the magnon
energies are independent of the wave number:
$\omega_k\approx\omega=E_Z$. 

The interaction $V=g\sum_i \myvec{s}_i\myvec{S}_i$ between
electron spins
$\myvec{s}_i=\frac12\sum_{\sigma\sigma'}c^\dagger_{i\sigma}
(\boldsymbol{\sigma})^{\phantom{\dagger}}_{\sigma\sigma'} c^{\phantom{\dagger}}_{i\sigma'}$ and
localized spins $\myvec{S}_i$ is transformed
using the HPT to $V=V^{(1)}+V^{(2)}+\Delta$, where
\begin{equation*}
  V^{(1)}=g\sqrt{\frac{S}{2N}}\sum_{kq}\left(b^\dagger_q
    c^\dagger\ikmqd 
    c\pd\iku+b\pd_q c^\dagger\ikpqu c\pd\ikd\right)
\end{equation*}
corresponds to spin flips of a conduction electron which involve the
emission or absorption of a magnon,
$V^{(2)}=\frac{g}{2N}\sum_{kqq'\sigma}\sigma b^\dagger_{q+q'}b\pd_q
c^\dagger_{k-q'\sigma}c\pd\iks$ 
implicates electron-magnon scattering, and $\Delta$ is a
spin-dependent energy shift which can be combined with
$\varepsilon\iks$ to form a new one-electron energy
$\bar\varepsilon\iks=-2t\cos(ka)+\frac12 \sigma (h_e-gS)$.

The wire is coupled to the leads by
$H_{\text{T}}=\sum_{\alpha lk\sigma}\left(t^\alpha_{\sigma}
  a^\dagger_{\alpha l\sigma}c\pd\iks+\text{H.c.}\right)$. 
Here, $\alpha\in\{\text{L},\text{R}\}$ labels the lead
and $l$ the electronic states with energies $\varepsilon_{\alpha
  l\sigma}$. The leads are assumed to be
noninteracting and to have a constant, but possibly spin-dependent,
density of states. This is reflected in the energy-independent
coupling function $\Gamma^\alpha\is=\Gamma^\alpha\is(E)=2\pi\sum_l
t^\alpha_{\sigma}t^{\alpha*}_{\sigma}\delta(E-\varepsilon_{\alpha
  l\sigma})$.

We assume that the occupation of magnon states is equal to the
equilibrium value
$n(\omega)=\left(\exp(\beta\omega)-1\right)^{-1}$, i.e., that the
coupling to an external spin bath causes magnon relaxation on
a time scale $\tau_{\text{M}}$ which is smaller than the average time
between two electron transmissions through the wire but larger than
the time needed to establish a coherent electron-magnon
state~\footnote{In terms of the energy scales involved,
  $\tau_{\text{M}}$ has to fulfill
  $|\varepsilon\iu-\varepsilon\id-gS-\omega|^{-1}\ll\tau_{\text{M}}\ll
  \Gamma_\sigma^{-1}$ to justify this approximation.}.

\sect{Method}The nonequilibrium Green function method proposed by
Keldysh~\cite{Keldysh} is used to calculate the current through the
wire which can be expressed in terms of the Green
functions~\cite{MeirWingreen}. We divide the electron self-energy into
two parts, $\Sigma=\Sigma_{\text{T}}+\Sigma_{\text{M}}$, where
$\Sigma_{\text{T}}$ is due to $H_{\text{T}}$. Its retarded/advanced
and Keldysh components are
$\Sigma_{\text{T}\sigma}^{\text{R,A}}=\mp\frac{i}{2}\Gamma\is$ and
$\Sigma_{\text{T}\sigma}^{\text{K}}(E)=i\sum_\alpha\Gamma^\alpha\is(2f_\alpha(E)-1)$,
respectively, where $\Gamma\is=\sum_\alpha\Gamma^\alpha\is$ and
$f_\alpha(E)$ is the Fermi function for lead $\alpha$, which has the
chemical potential $\mu_\alpha$.

The self-energy contribution $\Sigma_{\text{M}}$, which is due to the
electron-magnon interaction $V$, is calculated in
self-consistent Born approximation (SCBA), which corresponds to the
consideration of diagrams of the order $\mathcal{O}(g^2)$. We
evaluate $\Sigma_{\text{M}}$ using the free magnon Green functions which
do not depend on the magnon wave number due to the approximation
$\omega_k\approx\omega$. Therefore, $\Sigma_{\text{M}}$ is independent
of the electron wave number. With $M_\sigma (E)=\sum_k
|G^R_{k\sigma}(E)|^2$, its imaginary part is given by
\begin{equation}
  \text{Im}\,\Sigma_{\text{M}\sigma}^R(E+\sigma\omega)
  =-\frac{g^2S}{4N}M_{-\sigma}(E)
  \sum_\alpha \Gamma^\alpha_{-\sigma}
  f_\alpha^{-\sigma}(E),
  \label{eq:SelfEnergy}
\end{equation}
where $f^+_\alpha = f_\alpha$, $f^-_\alpha= 1-f_\alpha$. The real part
is obtained from the Kramers-Kronig relations. Terms which are
proportional to $n(\omega)$ have been omitted because $n(\omega)\ll1$
for the parameters chosen in the next section. Dyson's equation
and~\eqref{eq:SelfEnergy} are solved self-consistently using an
iterative procedure~\footnote{We remark that a non-self-consistent
  solution (which corresponds to stopping the procedure after the
  first iteration) yields qualitatively different results.}.

Richmond~\cite{Richmond} obtained the exact $\uparrow$-electron
self-energy for a single conduction electron in equilibrium by
considering a larger set of diagrams involving an arbitrary number of
electron-magnon scattering vertices between the emission and
absorption vertices. However, these diagrams would give rise to a
violation of charge conservation in the nonequilibrium situation which
is discussed here (i.e. the sum of the currents from the leads would
be nonzero: $I\iL+I\iR\neq0$). We therefore restrict our calculation
to the charge-conserving SCBA and disregard diagrams of higher order
than $\mathcal{O}(g^2)$~\footnote{We remark that the problem could be solved
if another set of more complex self-energy diagrams was taken into
consideration additionally. The discussion of a charge-conserving
approximation involving higher-order diagrams will be addressed in a
future publication, but does not change the qualitative effects
discussed here.}.

Electrons
which tunnel into the system from lead $\alpha$ with spin $\sigma$ can
either tunnel to lead $\alpha'$ with unchanged spin $\sigma$ (elastic
current) or flip their spin by emitting or absorbing a magnon and
leave the system with spin $-\sigma$ (inelastic current). The two
current contributions for lead $\alpha$ are
\begin{align*}
  I^{\text{el}}_\alpha=\frac{e}{h}&\int dE\,\sum_{\alpha'}\sum_\sigma
  T^{\text{el}}_{\alpha{\alpha'}\sigma}(E)
  \left(f_\alpha(E)-f_{\alpha'}(E)\right),\\ 
  I^{\text{inel}}_\alpha(\omega)=\frac{e}{h}&\int dE\,\sum_{\alpha'}
  \bigg(T^{\text{inel}}_{\alpha{\alpha'}}(E,\omega)f_\alpha(E)
  \big[1-f_{\alpha'}(E-\omega)\big]\nonumber\\
  &\quad\quad+T^{\text{inel}}_{{\alpha'}\alpha}(E,\omega)
  \big[1-f_\alpha(E-\omega)\big]f_{\alpha'}(E)\bigg),
\end{align*}
where we have again neglected terms~$\propto n(\omega)$. The
transmission coefficients are given by
\begin{eqnarray} 
T^{\text{el}}_{\alpha{\alpha'}\sigma}(E)&=&
\Gamma^\alpha\is\Gamma^{\alpha'}\is M_\sigma
(E),\\
T^{\text{inel}}_{\alpha{\alpha'}}(E,\omega)&=&\frac{g^2S}{2N}
\Gamma^\alpha\iu \Gamma^{\alpha'}_{\downarrow} 
M_{\uparrow}(E) M_{\downarrow}(E-\omega).
\label{eq:TransInelastic}
\end{eqnarray}
We consider both nonmagnetic and ferromagnetic
leads with polarization
$P_\alpha=(\Gamma^\alpha_\uparrow-\Gamma^\alpha_\downarrow)
/(\Gamma^\alpha_\uparrow+\Gamma^\alpha_\downarrow)\neq0$. 

\begin{figure}
  \includegraphics[width=\linewidth]{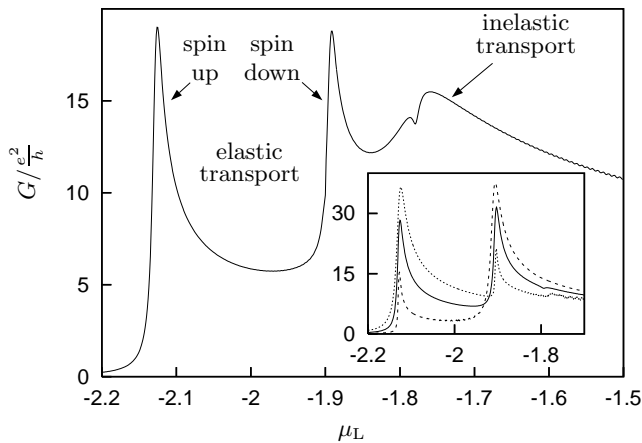}
  \caption{
    \label{fig:plot_N1000_g0_5}
    Differential conductance for a system with $N=1000$ sites and
    $S=1/2$. The parameters are $t=1$, $g=0.5$, $E_Z=h_e=0.1$, $k_BT=
    5\times10^{-4}$, $\sum_\sigma\Gamma^{\text{L,R}}_\sigma=10^{-2}$,
    $P\iL=+0.7$, $P\iR=-0.7$. The chemical potential $\mu\iR$ of the
    right lead is fixed at $-2.5$. Inset: Conductance for
    unpolarized leads (solid line) and ferromagnetic leads with
    parallel polarization and $P=+0.7$ (dotted line), $P=-0.7$ (dashed
    line).} 
\end{figure}

\sect{Results}The differential conductance for a large system with
$N=1000$ sites where the level spacing is smaller than the other
relevant energy scales is shown in
Fig.~\ref{fig:plot_N1000_g0_5}. Since the electron density in the wire
should be sufficiently low to justify the neglect of the Coulomb
interaction, we examine the voltage regime where only a small fraction
of the $\uparrow$- and $\downarrow$-electron states is partially
occupied. We fix $\mu\iR$ below the conduction band at $\mu\iR=-2.5$
and vary $\mu\iL$ (all energies are in units of $t$ which we set to
$t=1$).

For low bias voltages ($\mu\iL\lesssim-1.85$), elastic transport
processes (corresponding to resonances of $\big|G\ret\iks(E)\big|^2$) 
dominate.
One peak at $\mu\iL\approx-2.12$ for spin-up is due to the magnetic
polaron band, and the other at $\mu\iL\approx-1.89$ corresponds to
$\downarrow$-electron states. The shape of the peak
structures reflects the density of states in a one-dimensional system.
Inelastic processes superpose these structures. In the situation where
$P\iL=+0.7$ and $P\iR=-0.7$ (main plot of
Fig.~\ref{fig:plot_N1000_g0_5}), they dominate for higher voltages
($\mu\iL\gtrsim-1.85$) because this setup maximizes the product
$\Gamma\iu\hL\Gamma\id\hR$ which is proportional to the inelastic
transmission coefficient~\eqref{eq:TransInelastic}. The inset of
Fig.~\ref{fig:plot_N1000_g0_5} shows how the  different peaks can be
distinguished by changing the spin polarization of the leads. Here,
three conductance curves are shown for unpolarized leads and
ferromagnetic leads with parallel polarizations ($P\iL=P\iR=P$). The weight of the peak structure which is
related to elastic transport of electrons with spin $\sigma$ is
greatest if the leads are $\sigma$-polarized
($\sigma\in\{\uparrow,\downarrow\}$), just as one would expect. While
there is also an inelastic current contribution in these
configurations, it does not cause a clear signal in the differential
conductance. 

\begin{figure}
  \includegraphics[width=\linewidth]{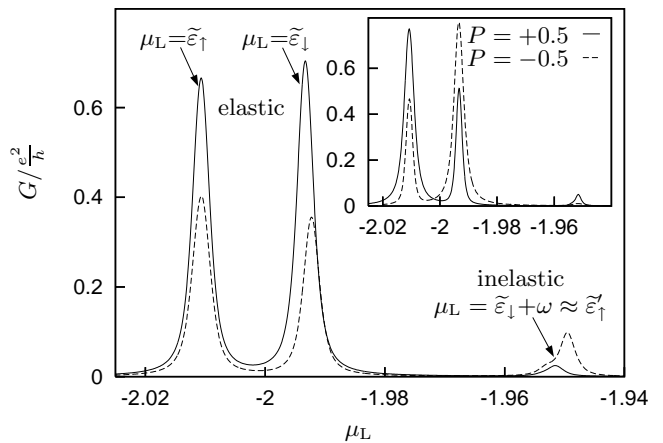}
  \caption{
    \label{fig:plot_N12_g0_1}
    Conductance for $N=12$, $S=1/2$. The parameters are $t=1$,
    $g=0.1$, $E_Z=h_e=0.04$, $k_BT=5\times 10^{-4}$,
    $\sum_\sigma\Gamma^{\text{L,R}}_\sigma=2\times10^{-3}$,
    $\mu\iR=-2.5$. Both nonmagnetic leads (solid line) and
    spin-polarized ferromagnetic leads with $P\iL=+0.7$, $P\iR=-0.7$
    (dashed line) are considered. Inset: $G$ for ferromagnetic leads
    with parallel polarization.}
\end{figure}
For a small system with $N=12$ sites, the discrete structure of the
energy spectrum can be identified in the differential
conductance if the level spacing is
larger than the energy scales $\Gamma$ and $k_BT$ which determine the
broadening of the conductance peaks. In the considered voltage regime,
only the lowest electronic states (wave number $k=0$, spin $\uparrow$
or $\downarrow$) are partially occupied and contribute to the
current. The electronic spin-up states are split mainly by the $q=0$
magnon into two states with energy $\euone$ and $\eutwo$
(corresponding to the magnetic polaron and the scattering state
in the continuum case, respectively). Futhermore, the electronic spin
down states are also renormalized to $\edone$ when the $\uparrow$-
electron states have a finite occupation probability.

Results for $g>0$ (antiferromagnetic local exchange coupling) are
presented in Fig.~\ref{fig:plot_N12_g0_1}. As in the situation
discussed above, conductance peaks which are due to elastic transport
processes coincide with resonances of the retarded Green functions. Here, the left
and right large peak occur at $\euone$ and $\edone$, the main
resonances of the $\uparrow$- and $\downarrow$-electron Green function (with wave
number $k=0$), respectively, and can be attributed to elastic
transport through the wire. On the other hand, the inelastic
transmission coefficient~\eqref{eq:TransInelastic} is proportional to
both $\big|G^{\text{R}}\iku(E)\big|^2$ and
$\big|G^{\text{R}}_{k\downarrow}(E-\omega)\big|^2$.
Therefore, inelastic transport processes contribute to the
differential conductance at $\mu\iL=\euone$ (where some weight is
added to the left large peak) and $\mu\iL=\edone+\omega$, the position
of the smaller peak. The dependence of the relative peak heights on
the lead polarizations is like in the large system discussed above.

Actually, one could expect another peak in
Fig.~\ref{fig:plot_N12_g0_1} because the $\uparrow$-electron Green function has a
second resonance at an energy $\eutwo$. However, it has in
general quite a small weight because the magnitude of the imaginary
part of the $\uparrow$-electron self-energy is rather large at
$\eutwo$, leading to a strong decay of the corresponding state and a
suppression of elastic transport. Moreover, $\eutwo$ is very close to
the energy $\edone+\omega$ where inelastic transport contributes to
the current. Therefore, elastic transport of
$\uparrow$ electrons is visible only at $\euone$ (energy of the
magnetic polaron), the small contribution at $\mu\iL=\eutwo$ is
absorbed in the inelastic peak. 

It should be noted that not only the peak heights but also the
positions depend on the polarizations of the
leads. Fig.~\ref{fig:plot_N12_g0_1} shows that the peak at
$\mu\iL=\edone$ and the inelastic peak are shifted to the right for
the configuration where the leads have antiparallel polarizations. The
reason is that a $\downarrow$ electron (or hole) can interact
with magnons only by flipping its spin and occupying an
$\uparrow$ state. Therefore, the $\downarrow$-electron self-energy
(and thus the position $\edone$ of the main resonance of the Green function)
depends on the occupation probability of $\uparrow$-electron
states. This probabilty is given by
$F\iu(E)\approx\sum_\alpha\frac{\Gamma^\alpha\iu}{\Gamma\iu}f_\alpha(E)$
for a state with energy $E$ and is affected by both the chemical
potentials and the polarizations of the leads.

\begin{figure}
  \includegraphics[width=\linewidth]{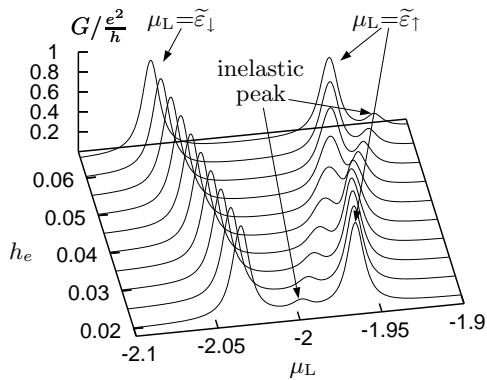}
  \caption{
    \label{fig:plot_N12_gm0_1_B}
    Conductance for $t=1$, $g=-0.1$, $k_BT=10^{-3}$,
    $\Gamma^{\text{L,R}}_\sigma=5.5\times 10^{-3}$, $\mu\iR=-2.5$. The
    Zeeman splittings $h_e$  for conduction electrons and $E_Z=2h_e$
    for localised spins are different for each curve.}
\end{figure}

Conductance curves for ferromagnetic local exchange coupling ($g<0$)
and different magnetic fields are shown in
Fig.~\ref{fig:plot_N12_gm0_1_B}. The Zeeman splitting  $E_Z$ of the
localized spins is chosen to be twice as large as the splitting $h_e$
of the conduction electrons. Therefore, not only the peak positions
but also the general structure of the conductance curve change if the
magnetic field is varied. For small ($h_e\approx0.02$) and large
($h_e\approx0.06$) fields, the situation is comparable to
Fig.~\ref{fig:plot_N12_g0_1}: There are two large ``elastic'' peaks at
$\mu\iL=\euone$ and $\mu\iL=\edone$ and a small ``inelastic'' peak at
$\mu\iL=\edone+\omega$. These peaks move with  different 'velocities'
if the field is increased: The positions of the large peaks, i.e., of
the main resonances of the retarded Green functions, change with the
conduction electron Zeeman energy $\pm h_e/2$, but the position of the
inelastic peak changes like $-h_e/2+E_Z=3/2 h_e$ because of our choice
$E_Z=2h_e$. One could expect that the inelastic and the right elastic
peak overlap and form a single resonance for intermediate fields
($h_e\approx0.045$), but this is not the case. The corresponding
conductance curve rather reveals two peaks of comparable height. These
arise from two resonances of the $\uparrow$-electron Green function
which have approximately equal weight for this particular set of
parameters. This means that the decoherence rates are equal for the
states corresponding to the energies $\euone$ and $\eutwo$, in
contrast to the situation in Fig.~\ref{fig:plot_N12_g0_1}. Elastic
transport of $\uparrow$ electrons thus generates a double-peak
structure in the differential conductance which is superposed by a small
inelastic transport contribution.

We remark that without coupling to the spin chain ($g=0$),
there would be only elastic transport through the
wire. The effects discussed here, i.e., inelastic transport, a
dependence of peak positions on the lead polarization, and magnetic
field-dependent decoherence rates of the states involved in
transport, would not occur.

\sect{Summary}We presented a self-consistent diagrammatic approach
within the Keldysh formalism to calculate the nonequilibrium current
through a mesoscopic quantum wire coupled to a ferromagnetic spin
chain. We proposed a way to detect the coherent superposition of
electronic and magnon states, the so-called magnetic polaron.
It shows up as a high (i.e. coherent) signal in the differential
conductance and can be tuned by external magnetic fields and the
spin polarization in the leads. In this way we have shown that the
interaction between electrons and magnons (which usually leads to
unwanted relaxation and dephasing of the electron spin) can be used
for the creation of a phase-coherent quantum state. We
expect that this work will stimulate further theoretical and
experimental investigations of the magnetic polaron in the field
of mesoscopic systems. 

\begin{acknowledgments}
The authors would like to thank S. Jakobs and
J. K\"onig for helpful discussions. This work has been supported by the
VW Foundation and the Forschungszentrum J\"ulich (via the virtual 
institute IFMIT).
\end{acknowledgments}

\end{document}